\documentstyle[aps,epsf,twocolumn,prl]{revtex}

\newcommand{\beq}{\begin{equation}} \newcommand{\eeq}{\end{equation}}   
   
\begin{document} 
 \author{ Alexei V. Tkachenko \\
\it Bell Labs, Lucent Technologies, 600-700 Mountain Ave., Rm 1D329, Murray Hill, NJ 07974 \\
E-mail: alexei@bell-labs.com }
\title{\bf With a little help of DNA:
Morphological diversity of colloidal self--assembly  }  
\maketitle
\begin{abstract}
We study theoretically a  binary  system  in which an  attraction  of  unlike particles is combined with   a type--independent  soft core repulsion.  The    possible   experimental implementation  of the system  is  a  mixture of DNA--covered colloids, in which both the repulsion and the attraction may be   induced by   DNA solution.   The system is shown to  exhibit surprisingly diverse and unusual morphologies. Among them are  the diamond lattice and  the  membrane phase with in--plane square  order, a striking example of {\sl spontaneous compactification}. 
 \end{abstract}

Self--assembly in colloidal systems has  attracted  a lot of interest      as an  experimental tool  for study  of   crystallization and glassiness \cite{colloids}. It also has a  great potential for the  submicron technology, especially for fabrication of photonic band gap materials\cite{photonic}.
The progress in these directions   is considerably limited by the lack of diversity of the crystalline   morphologies achievable by self-assembly. Typically,   a monodisperse colloidal system  crystallizes into a  close--packed structure (FCC, or another stacking of  hexagonal layers)\cite{colloids},\cite{chai}.

In this paper, we propose  a   system which combines   relatively  simple  interactions with a  rich and unexpected phase behavior. It  is inspired by recent experimental demonstration  of  DNA--assisted  self--assembly of nanoparticles \cite{mirkin1,mirkin2}. The key elements of that scheme are submicron spheres (e.g.  golden, silica, or other) covered with  short single--stranded DNA ``markers''.  The  marker sequence  determines the particle type (there may be many markers per particle, but their  sequences must be the same).   One can now introduce  type--dependent interactions between the  particles by  adding ``linker'' DNA molecules, whose ends are complementary to the corresponding markers. These interactions are  very selective, reversible and tunable.

We start our  discussion with  a  generic model  in which the physical origin and   details of the inter--particle interactions are  largely ignored. Let us  consider a binary system of  spheres, of the  same diameter $d$,   repelling each other with a  soft--core   potential, $U(r)$.   In addition, the  unlike particles  may   stick to each other with  binding energy  $-E$. We will study the phase behavior of the system for various values of  two  parameters: the aspect ratio, $d/\xi$ ($\xi$ is the range of potential $U(r)$), and the relative strength of the attraction, $E/U_0$ (here $U_0\equiv U(d)$). Later in the paper,  we discuss how this system can be implemented experimentally. Our specific proposal is to use DNA   to introduce type-dependent attraction, and polymeric (possibly, also DNA) brush to induce  the  repulsive potential.  

The non-trivial phase behavior of the discussed system is a result of interplay between the  adhesive energy and the soft--core  repulsion. We will consider only structures with $1:1$ composition. Let $r_k$ be a distance to the $k$--th nearest neighbor of a particle in a given structure ($r_1\equiv d$), let $Z_k$ be the average  number of such neighbors, and  $Z$ be the average number of cohesive contacts per particle (coordination number).   If the entropic effects are neglected, the energy per particle (i.e. the average chemical potential of A and B particles)  is

\begin{equation}\mu= \frac 12\left(- Z E +\sum_{k=2}^\infty Z_k U(r_k)\right) \simeq \frac 12\left(- Z E+Z_2 U(r_2)\right)\end{equation}

Here, we have neglected the contribution  from the particles beyond    the  second nearest neighbors, which is a reasonable  approximation  for $d/\xi\gg 1$. In order to  assure its validity, we perform a posteriori check of the effect of the  higher order corrections on our  results. It is straightforward to use the above equation to identify the phase  boundary between two different structures. Since the chemical potential should be continuous at the transition ,  one can express the critical  value of adhesive energy $E$  in terms of the  geometrical parameters ($Z$, $Z_2$, and  $r_2$) of the two phases

\begin{equation} \label{trans}
E\simeq \frac{Z'_2U(r'_2)-Z_2 U(r_2)}{Z'-Z}\end{equation}

The task of identifying  all plausible morphologies of our system is clearly more challenging than comparing them energetically.    There is hardly any systematic way of doing this, which would go beyond an educated guess. Nevertheless, one can considerably  limit the search by making a number of   assumptions based on general principles \cite{Henley}:
(i) In order to avoid direct contacts of the same--type particles, the structures should be bipartal, i.e. they  should consist of two sub--lattices corresponding to the two types of particles, so that all the  nearest neighbors were of opposite types. 
(ii) All the  nearest neighbors should  have the same bond length.  This is needed to take advantage of the cohesive energy, as long as we model the particles as rigid sticky spheres.
(iii) The structure is likely to possess a high symmetry: we consider only crystalline morphologies, with all equivalent  sites.  The violations of this criterium are fairly rare (e.g. quasicrystals).

In  the table below, the candidate phases satisfying the above principles are classified accordingly to their coordination numbers and the space dimensionality.
\begin{center}
\begin{tabular}{|c|c|c|}
\hline
Z & 2D  & 3D \\
\hline 3  & Honeycomb & Gyroid \\
\hline 4 & Square ({\bf SQ}) & Diamond ({\bf D})\\
\hline 5 & & Honeycomb Stacking ({\bf HS}) \\
\hline 6 & & Simple Cubic  ({\bf SC})\\
\hline 8 & & Body-Centered Cubic ({\bf BCC})\\
\hline
\end{tabular}
\end{center}

  Note that in the limit of large aspect ratio, among various phases with the same $Z$, the one with the largest $r_2$  is energetically preferred, independently on the number of the second nearest neighbors ($Z_2$). Likewise, since $r_2$  goes down with the increase of coordination number, $Z$, the transition value of $E$, given by  Eq. (\ref{trans}), asymptotically reaches value $ Z_2U(r_2)$, determined by the parameters ($Z_2$, $r_2$) of the higher-$Z$ phase.  In other words, the higher $Z$, the higher   is the parameter $E(Z)$ of the transition between $Z$-- and $Z+1$--coordinated phases. 
 One can conclude that  the sequence of the phases in the large aspect ratio regime is generic and should be independent of  the particular choice of the repulsive  potential: BCC($Z=8$)--SC($Z=6$)--Honeycomb Stacking ($Z=5$)--Diamond($Z=4$). The possibility of the self--assembly of the  diamond lattice  is especially exciting, because of its potential as a     photonic bad gap structure \cite {photonic}.

The energetic difference between the two $Z=3$ phases in the above table  is only due to interactions of the third--nearest neighbors, which makes it irrelevant for a realistic situation. Unless the system is confined to   2D, this  suggests that the observed morphology with $Z=3$ will typically be a  disordered one.  There is no true phase transitions for ($Z<3$) because of the low dimensionality of the dominant structures.

\begin{figure}[h]
a)
\centerline{ \epsfxsize 8truecm \epsfbox {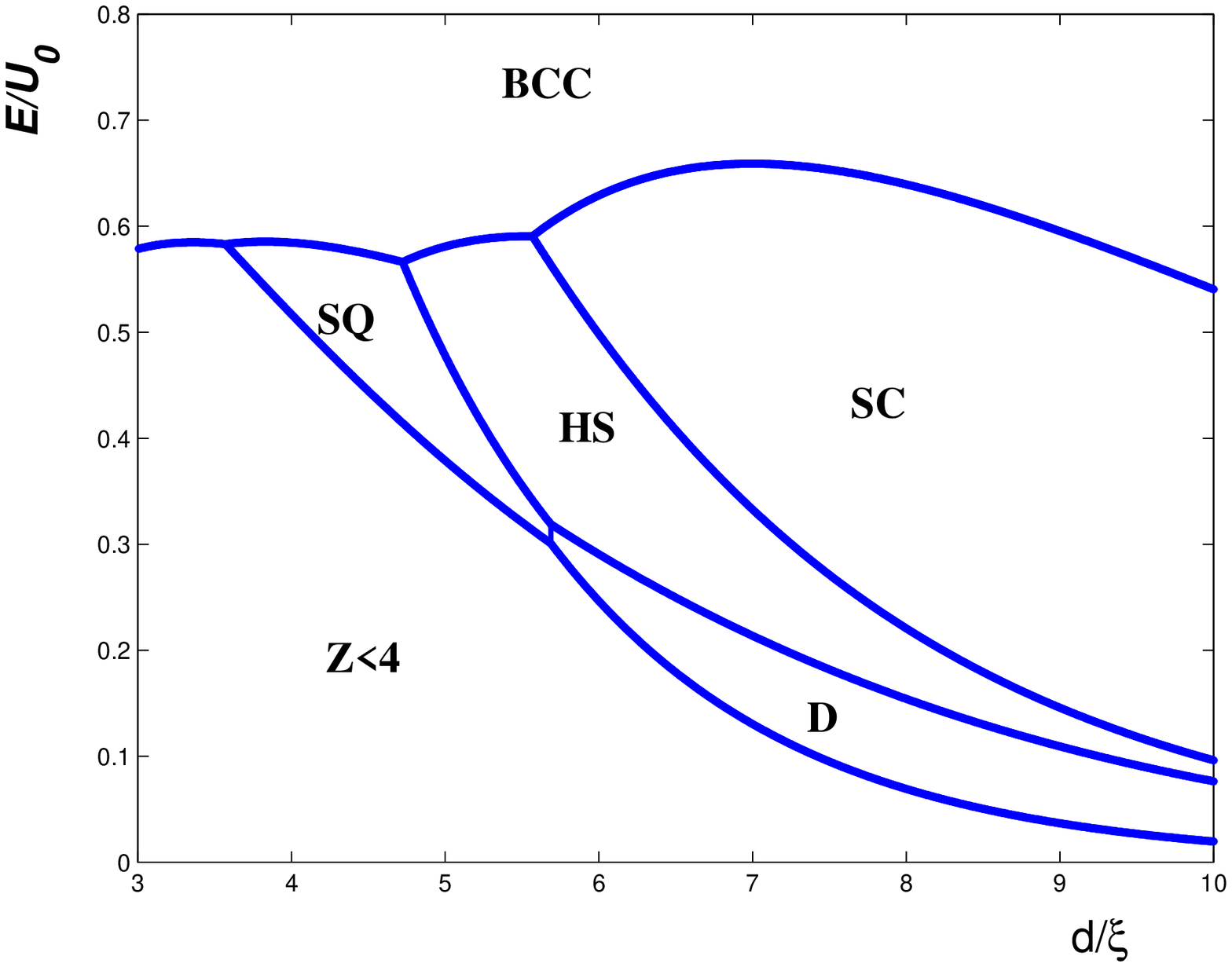} } 
b)
\centerline{ \epsfxsize 8truecm \epsfbox {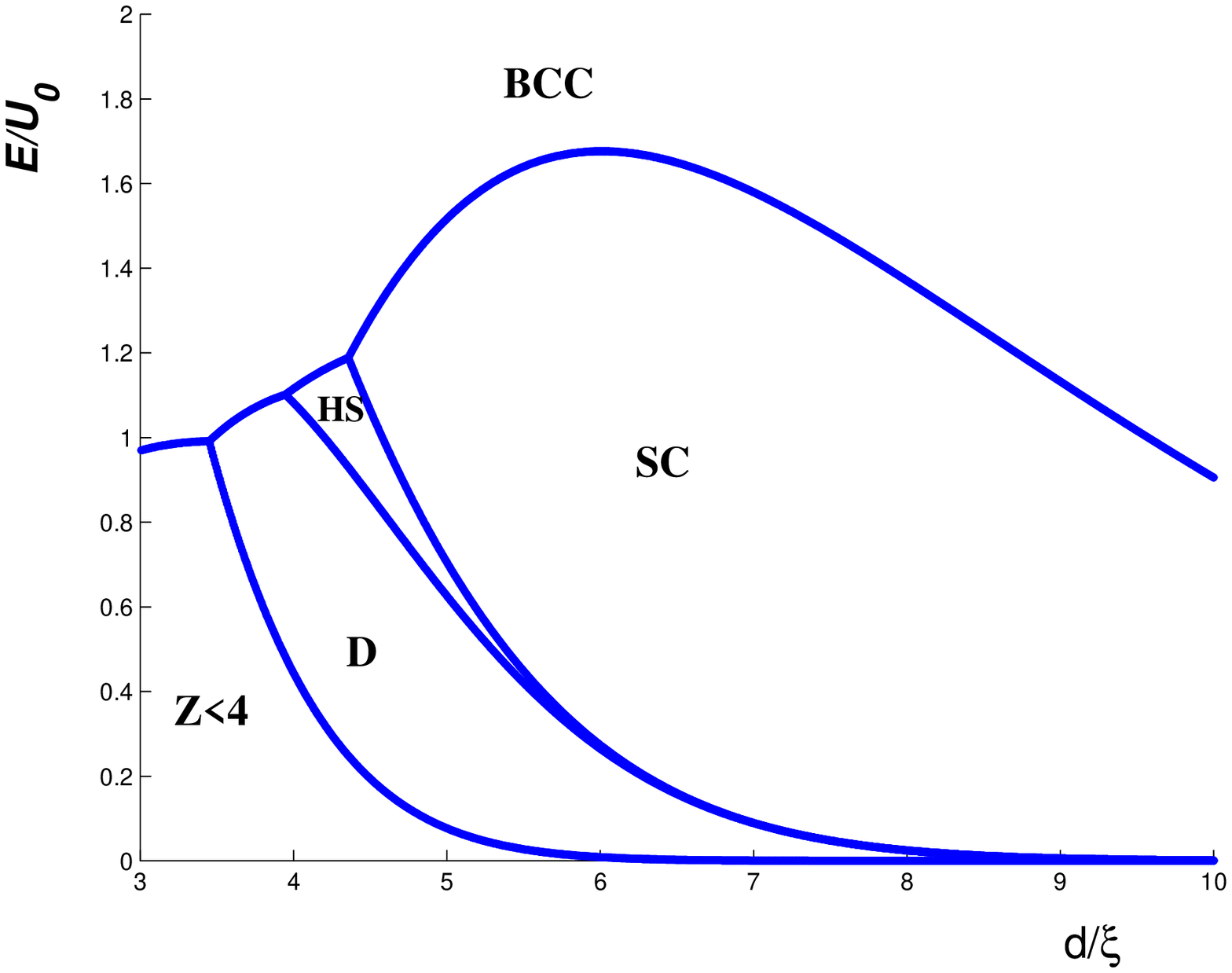} }
 \caption{Calculated phase diagram of the system for exponential (a) and Gaussian (b) forms of the repulsive potential $U(r)$. }
\label{phase_diag}   
\end{figure}

As the aspect ratio $D/\xi$ is being decreased, one can expect two types of events: (i) "squeezing out" of  certain phases by their neighbors; (ii)  replacingone   structure with  another without changing  $Z$. As one can see from Figure \ref{phase_diag}(a), the phase diagram obtained for exponential potential,  $U(r)=U_0\exp(-(r-d)/\xi)$ provides examples of the both kinds. One of them  is particularly striking: at certain aspect ratio, the system undergoes a transition from 3D diamond lattice to 2D membrane with in-plane square order. Even though this order is known to be strongly affected by long-range fluctuations \cite{chai}, the corresponding  Landau-Pierls effect does not have any divergent contributions to the chemical potential of the 2D phase.  One may  refer to  $\bf D-SQ$ transition as {\em spontaneous compactification}. Although the found 2D phase is somewhat similar to lipid membranes, it is built by particles with isotropic  effective  interactions (unlike lipids).  The same holds for the diamond lattice: unlike the diamond structures in nature, in our case it is due to the competition of relatively simple {\em isotropic}   potentials!      

As it was argued, the major features of the obtained phase diagram are fairly independent of the choice of the form of the inter--particle potential. To  check this, we have calculated  the phase diagram for two   types of interactions, exponential and Gaussian, $U(r)= \exp(-(r-d)^2/2\xi^2)$ (see Figure \ref{phase_diag}). The major difference is that the square lattice ({\bf SQ}) completely disappears in the Gaussian case. In addition  the diamond lattice ({\bf D})  significantly expands at the expense of {\bf SQ} and {\bf HS} phases. This trend appears to be quite general: the balance between the two competing $Z=4$ phases,  {\bf D} and {\bf SQ}, shifts towards Diamond for potentials with a  super--exponential decay, while the region of stability of  {\bf SQ}  expands  for sub-exponential $U(r)$  (such as  power laws, stretched exponentials, or Yukawa interaction).

We now proceed with the   discussion  of a plausible experimental implementation of the proposed system. 
 As we have already mentioned,  type--dependent  "DNA bridging"  of colloidal particles is an appealing way to introduce the   AB attraction. As to the repulsive interactions, there are several candidates in the  colloidal science. Unfortunately, using of electrostatic repulsion  in our case  is  problematic:  in order for it   to be relevant, the  salt concentration needs to be much lower than at the physiological conditions. As a result,  there would be a high  electrostatic barrier for   DNA duplex formation.

An alternative way to introduce the soft--core repulsion is through  steric interactions  of  polymer--covered particles. Here  we focus on  a  particular scenario in which the repulsion is due to DNA molecules with only one ``sticky end'' (complementary  to  either A or B markers). Unlike linker DNA,  these one--arm molecules do not result in bridging and play a role of a buffer.  We will assume that both the buffer--DNA and the linkers are double stranded, with the exception of the short terminal segments. The strength of the interaction of these sticky ends with the complementary markers can be characterized by  DNA concentration  $c_0$, at which the condensation would  occurs (i.e. when the chemical potential of an  adsorbed chain would become  equal to that in the solution). If   the actual concentration of the buffer--DNA, $c$, is much lower than  $c_0$, the number of the adsorbed chains per particle is $N=N_{\max}c/c^{(0)}$. Here $N_{\max}$ is the total number of markers per particle. For the sake of simplicity,  we assume that $N^{\max}$ and $N$ are the same for A and B particles, i.e.  $c_A/c_B=c^{(0)}_A/c^{(0)}_B$.

\begin{figure}[h]
\centerline{ \epsfxsize 8truecm \epsfbox {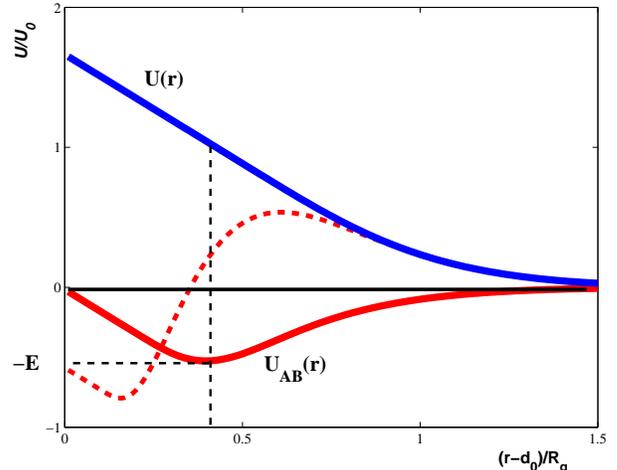} } 
 \caption{Repulsive, $U(r)$, and attractive , $U_{AB}(r)$ potentials induced by DNA-particle interactions. Solid lines correspond to $R'_g/R_g=1$. Note the barrier in the attractive potential for $R'_g/R_g=0.5$ (dashed line).}
\label{potentials}   
\end{figure}

Note that we have totally neglected the excluded volume interactions between the DNA molecules, which might seriously decrease the coverage compared to our result.  
The reason for this is twofold. First, the optimal coverage (discussed below) corresponds to a moderate overlap between the adsorbed chains. Second,  the high rigidity of a double-strand DNA molecule results in  the excluded volume effects being negligible within a single chain as long as it remains shorter than   several thousands  persistence lengths, $l_p\simeq 50\  nm$. Thus, the adsorbed DNA molecules  can be treated as phantom Gaussian chains, which only interact with hard surfaces of the particles, but not with each other. Since the adsorption is reversible, the confined chains are being  ``squeezed out'' when the gap between two particles, $r-d_0$ becomes comparable to the gyration radius of the DNA chain, $R_g$. The corresponding energetic penalty    can be calculated in the spirit of the  Deryagin approximation\cite{isr} (from now on, we  distinguish     bare particle diameter, $d_0$,   and  effective one, $d$):
\begin{equation} 
\label{repulsion}
U(r)\simeq 2N kT \frac{R_g}{d_0}\int_{(r-d_0)/R_g}^{\infty}\left[1-\exp\left(-\frac{W(\Delta)}{kT}\right)\right]{\rm d}\Delta \end{equation}
Here $W(\Delta)$ is the free energy penalty for the  confinement of a Gaussian polymeric chain between two walls at separation $ R_g \Delta$. It can be obtained by using the Schr\"odinger--like description of the ideal polymer\cite{polymer1,polymer2}:  
\begin{equation}\frac{W(\Delta)}{kT}=\log\left(\frac{\sqrt{2/3}}{\Delta}\sum_{n=0}^{\infty}\exp\left[ -\frac{1}{6}\left(\frac{\pi (1+2n)}{\Delta}\right)^2 \right]\right) \end{equation}
The resulting repulsive potential is shown on Figure \ref{potentials}. 

 The attractive potential induced by the linker--DNA can be calculated in a very similar manner. An important differences are  that one has to take into account the elastic energy of a stretched linker, $W_{el}(\Delta)=3kT\Delta^2/2$, and  that the condensation concentration of the chains with two sticky ends  is given by $c_{AB}^{(0)}=c_A^{(0)}c_B^{(0)} R_g'^3$. Note that in a general case  the gyration radius of a linker--DNA, $R'_g$, may be  different from $R_g$.

\begin{eqnarray} 
\lefteqn{U_{AB}(r)\simeq U(r) - \frac{2N^2_{\max} kT}{3} \frac{c_{AB}}{c_{AB}^{(0)}}\left(\frac{R_g}{d_0}\right)^3 \times }&
 \\
\ \   &  {\int_{(r-d_0)/R_g}^{\infty}\exp\left[-\frac{W(\Delta)+W_{el}(\Delta)}{kT}\right]d\Delta} \nonumber 
\end{eqnarray}

The relative strength of the attraction and the repulsion is controlled by the ratio of the concentrations of the buffer--DNA  and linkers:

\begin{equation}
\frac{E}{U_0}=\frac{c_{AB}}{C_{AB}}-1
\end{equation}

Here $C_{AB}$  is  the linker concentration  at which $E$ vanishes. For the case  of  $R'_g=R_g$, 

\begin{equation}\label{alpha}
C_{AB} \simeq \frac{3 c^{(0)}_A c^{(0)}_B  R_g^5}{N_{\max} d^2}
\end{equation}

The absolute value of the energy scale in the problem is also controllable by concentration: according to  Eq. (\ref{repulsion}),

\begin{equation}\label{U0}
 U_0\simeq N kT\frac{R_g}{d}
\end{equation}
 This scale should be of order of several $kT$ to ensure that the phase diagram is not affected by the thermal fluctuations,  and that the escape time from the potential well of depth $E\equiv -\min\left(U_{AB}(r)\right)$ is not too long. By comparing Eqs. (\ref{alpha}) and (\ref{U0}), we conclude that $C_{AB}\simeq c_A c_B R_g^6/d^3$. 

As one might expect, the tail of repulsive potential $U(r)$ is well described by a Gaussian with the characteristic length scale $\xi\simeq R_g$: 
\begin{equation}
U(r)\sim \exp\left[-\frac{3}{2}\left(\frac {r-d}{R_g} + \beta \right)^2\right]. 
\end{equation}
Here d is the effective diameter which is determined by the position of the minimum of $U_{AB}(r)$, and the bias $\beta \simeq 0.15+(d-d_0)/R_g$. For the case $R'_g/R_g=1$, shown on Figure \ref{potentials}, $\beta \simeq 0.6$. Thus, the corresponding phase diagram should be somewhere  halfway between the Gaussian and exponential ones. By increasing  the ratio $R'_g/R_g$ one can  move  the system more towards  exponential regime, because the position of the minimum changes roughly linearly with the radius of the linker DNA. However, the dynamic range of $R'_g/R_g$ is rather limited: the  long linkers would  result in the additional   attraction beyond the nearest neighbors, which would violate  our initial assumptions. 

In the opposite regime,   $R'_g<R_g$, the particles need to overcome a significant energetic barrier before they start feeling the attraction (see  Figure \ref{potentials}). The existence of this barrier is the major reason why we suggest to use linkers of  at least several persistence lengths. In this case, the above  Gaussian description  may be applied. As we have shown, when   the gyration radius of the linkers matches  the scale of the repulsive potential, the particles may  create a bound state without the need of overcoming any  barrier. Thus, if the linkers are double--strand DNA molecules, the minimal value of $\xi_{\min} \sim l_p \sim 100 nm$.

The fundamental time scale of the problem is set by the lifetime of the $AB$ contact, which can be estimated as
\begin{equation}
\label{time}
\tau \simeq \frac{\eta d R'^2_g}{kT}\exp\left(\frac{E}{kT}\right).
\end{equation}
Here $\eta$ is the solvent viscosity, and we have assumed that the trial frequency is limited by the particle diffusion rather than by the departure rate of a linker--DNA. Since this desorption rate  can be  estimated as $kTc^{(0)}_{AB}/\eta$, our  regime  requires $ c^{(0)}_{AB} >R'^{-3}_g$.  This  corresponds to a moderate  marker--linker affinity, $\epsilon/kT < 3\log (R_g/a)\simeq 20$ (here $a\sim 0.1 nm$ is the microscopic scale defined by the rigidity of hydrogen bonding between complementary  DNA threads).

Since the relaxation time  (\ref{time}) grows fast with the liner scales of the problem, the optimal regime corresponds to minimal value of $R_g\simeq \xi_{\min}\simeq .1\mu m $, i.e  $d\sim 1 \mu m$. Within the assumption that the DNA--based key--lock complex would  remain functional up to the coverage of 1 marker molecule  per $\sim 10 nm^2$, $N_{\max} \sim 10^4$. According to Eqs. (\ref{alpha}-\ref{U0}), this yields $c_A/c_A^{(0)}=c_B/c_B^{(0)}\sim 10^{-3}$, and   $c_{AB}/c_{AB}^{(0)}\sim 10^{-5}$. The fundamental time scale (\ref{time}) at the optimal regime   is less than a  minute. However, the true relaxation time of the system is determined by slow  aggregation  and growth processes. Hopefully, they both can be facilitated by  using a  commensurate substrate.

Now that we have identified the optimal regime and interpreted the control parameters of the   phase diagram in terms of the experimental ones, we should also determine how damaging is disorder for the expected behavior. The effect of the particle  polydispersity  is likely to be similar to  the  one in conventional colloidal self--assembly. The new effect is that the interactions  are due to DNA chains, whose number per contact is of order of $1$. Nevertheless,  because of the reversibility of DNA--particle contact, the time--averaged interaction free energy varies only due to   the discreetness of the markers. The corresponding disoerder  can be estimated as $\delta E/U_0\simeq R_g\sqrt{N_{\max}}/d< 0.1$. As one can see from Figure \ref{phase_diag}, such a  variation of $E/U_0$ is tolerated in this    problem, especially for  the Gaussian potential.

{\bf Acknowledgement} The author is grateful to P. Wiltzius, C. Henley, B. Shraiman, and Z. Chen for the useful discussions and valuable information.


\begin{thebibliography}{99} 


\bibitem{colloids} P. Pieranski, {\sl Phys. Rev. Lett.}, {\bf  45}, 569 (1980);A. E. Larsen and D. G. Grier {\sl Phys. Rev. Lett.} {\bf 76}, 3862 (1996);   J. X. Zhu , M. Li, R. Rogers, W. Meyer, R. H. Ottewill, W. B. Russell, P. M. Chaikin, {\sl Nature} {\bf  387},  883 (1997).
\bibitem{photonic}  S. Fan, P. R. Villeneuve, and J. D. Joannopoulos, {\sl Phys. Rev. B}, {\bf 54},  11245 (1996);   D. F. Sievenpiper, M. E. Sickmiller, and E. Yablonovitch,  {\sl Phys. Rev. Lett.}{\bf 76},  2480 (1996).
\bibitem{chai} P. M. Chaikin and T. C. Lubensky, {\sl Principles of Condensed Matter Physics}, Cambridge Univ. Press., Camebridge, UK (2000).
\bibitem{mirkin1} C. A. Mirkin, R. L. Letsinger, R. C. Mucic,  J. J. Storhoff, {\sl  Nature}, {\bf  382}, 607 (1996); J. J. Storhoff, ; A.
A. Lazarides, R. C. Mucic, C. A. Mirkin, R. L. Letsinger,  G. C. Schatz, {\sl J. Am. Chem. Soc.} {\bf 122}, 4640 (2000).
\bibitem{mirkin2} J. J. Storhoff and  C. A. Mirkin,   {\sl Chem. Rev.},{\bf 99}, 1849 (1999).
\bibitem{Henley} W.-J. Zhu and C. L. Henley, {\bf Europhys. Lett.} {\bf 51}, 133 (2000);   W.-J. Zhu and C. L. Henley, Preprint (2001).
\bibitem{isr} J. N. Israelashvili, {Intermolecular and Surface Forces} (Academic Press, London, 1985)
\bibitem{polymer1}P.-G. de Gennes, {\sl Scaling Concepts of Polymer Physics} (Cornell University Press,   1979).
\bibitem{polymer2} A. Yu. Grosberg and A. R.  Khokhlov, {\sl Statistical Physics of Macromolecules} (AIP, New York, 1994).

\end{thebibliography}
\end{document}